\documentclass[floats, aps, prb, twocolumn, longbibliography, superscriptaddress]{revtex4-2}

\usepackage{url}
\usepackage{bm}
\usepackage{graphicx}
\usepackage{amsmath}
\usepackage{amstext}
\usepackage{amssymb}
\usepackage{amsfonts}
\usepackage{amsbsy}
\usepackage{verbatim}
\usepackage{xcolor}
\usepackage[colorlinks=true, urlcolor=blue, linkcolor=blue, citecolor=blue, pdftex]{hyperref}
\usepackage{multirow}
\usepackage{floatrow}
\usepackage{float}
\usepackage{gensymb}
\usepackage{textcomp}
\usepackage{enumitem}
\usepackage[version=4]{mhchem}
\usepackage{afterpage}
\usepackage{pbox}
\usepackage{makecell}
\usepackage{array,booktabs}
\usepackage{dsfont}
\usepackage{siunitx}
\usepackage[utf8]{inputenc}
\usepackage{braket}
\usepackage[normalem]{ulem}
\usepackage[version=4]{mhchem}
\usepackage[T1]{fontenc}
\usepackage[table,xcdraw,dvipsnames]{xcolor}
\usepackage{makecell}
\usepackage[normalem]{ulem}

\begin{document}

\title{Pressure-induced desalting of NaCl-bearing ice VII: A compositional filter in water-rich planets}

\author{Alasdair Nicholls}
\email{alasdair.nicholls@epfl.ch}
\affiliation{Laboratory for Quantum Magnetism, Institute of Physics, \'Ecole Polytechnique F\'ed\'erale de Lausanne (EPFL), CH-1015 Lausanne, Switzerland}
\affiliation{Institut de Min\'eralogie, de Physique des Mat\'eriaux et de Cosmochimie (IMPMC), CNRS UMR7590, Sorbonne Universit\'e, Paris, 75252, France}
 
\author{Tomasz Por\k{e}ba}
\affiliation{Laboratory for Quantum Magnetism, Institute of Physics, \'Ecole Polytechnique F\'ed\'erale de Lausanne (EPFL), CH-1015 Lausanne, Switzerland}
\affiliation{Paul Scherrer Institute (PSI), CH-5232 Villigen PSI, Switzerland}
 
\author{Xi Zhu}
\affiliation{Universit\'e Paris Cit\'e, Institut de Physique du Globe de Paris (IPGP),  CNRS, Paris, 75005, France}
 
\author{Richard Gaal}
\affiliation{Laboratory for Quantum Magnetism, Institute of Physics, \'Ecole Polytechnique F\'ed\'erale de Lausanne (EPFL), CH-1015 Lausanne, Switzerland}

\author{Stefan Klotz}
\affiliation{Institut de Min\'eralogie, de Physique des Mat\'eriaux et de Cosmochimie (IMPMC), CNRS UMR7590, Sorbonne Universit\'e, Paris, 75252, France}
 
\author{Gabriel Tobie}
\affiliation{Laboratoire de Plan\'etologie et G\'eosciences, UMR 6112, Nantes Universit\'e, 2 Rue de la Houssini\`ere, Nantes 44322, France}
 
 \author{Henrik M. R{\o}nnow}
\affiliation{Laboratory for Quantum Magnetism, Institute of Physics, \'Ecole Polytechnique F\'ed\'erale de Lausanne (EPFL), CH-1015 Lausanne, Switzerland}

\author{Razvan Caracas}
\affiliation{Universit\'e Paris Cit\'e, Institut de Physique du Globe de Paris (IPGP),  CNRS, Paris, 75005, France}

\author{Livia Eleonora Bove}
\email{livia.bove@upmc.fr}
\affiliation{Laboratory for Quantum Magnetism, Institute of Physics, \'Ecole Polytechnique F\'ed\'erale de Lausanne (EPFL), CH-1015 Lausanne, Switzerland}
\affiliation{Institut de Min\'eralogie, de Physique des Mat\'eriaux et de Cosmochimie (IMPMC), CNRS UMR7590, Sorbonne Universit\'e, Paris, 75252, France}
\affiliation{Dipartimento di Fisica, Sapienza Universit\`a di Roma, Piazzale Aldo Moro 5, Roma, 00185, Italy}

\date{\today}
 
\begin{abstract}
Sodium chloride is a major component of extraterrestrial oceans, but it is generally assumed to be excluded from the high-pressure ice mantles of water-rich planets. Here we show that a  concentrated \ce{NaCl} solution, when quenched in an amorphous state before compression, crystallises in a metastable \ce{NaCl}-bearing ice VII phase. Synchrotron x-ray diffraction and neutron diffraction reveals an expanded and strongly disordered ice VII lattice with no detectable crystalline NaCl. \textit{Ab initio} calculations reproduce lattice expansion by substituting water molecules for Na$^+$ and Cl$^-$ ions and show that the structural response increases strongly with salt concentration. Upon decompression at $300$\,K, the phase expels crystalline \ce{NaCl} at $5.1 \pm 0.3$\,GPa and does not reform upon recompression, demonstrating that its formation is controlled by the thermodynamic pathway. Despite an $8.5$\,\% larger unit-cell volume, \ce{NaCl}-bearing ice VII has a density close to that of pure ice VII and is substantially less dense than an equivalent mechanical mixture of ice and halite. These properties suggest that salty ice may be transported through deep convecting mantles but release salt near the $5$\,GPa boundary, where the denser products could inhibit further upward transport. Pathway-dependent salt trapping may therefore provide a previously unrecognised control on chemical exchange between rocky interiors and subsurface oceans.

\end{abstract}
 
\maketitle
 
\section{Introduction}

Most models of icy-moon and water-rich-planet interiors start from a simple assumption: as high-pressure ice crystallises at depth, dissolved salts are pushed out into the liquid above. Here we show that this assumption fails for ice VII. Using a `quench-and-compress' synthesis route, we demonstrate that NaCl can be extensively incorporated into the ice VII lattice as a genuine substitutional solid solution, with the local composition of the expanded phrase estimated to reach up to $33.1$\,wt\%. The subsequent consequences of this new phase affect the understanding of how salt is distributed and how it moves inside icy worlds.
 
The interiors of icy moons and water-rich planets are dominated by high-pressure phases of water ice, which control their density structure, internal stratification, and long-term thermal evolution \cite{Journaux2020, Journaux2022, Lichtenberg2025}. Whether salts are retained within these deep ice phases or expelled into subsurface oceans is one of the least constrained parameters in current interior models, yet it has an outsized effect on ocean salinity, ice-shell density, and ultimately on habitability. Sodium chloride (NaCl) is expected to be among the most abundant electrolytes in extraterrestrial oceans: it has been detected in NaCl-rich plume grains from Enceladus \cite{Hansen2006, Porco2006, Postberg2011, Postberg2009}, inferred spectroscopically on Europa and Ganymede \cite{Trumbo2019, Ligier2016, Tosi2023, Ligier2019}, and predicted in the geochemical models of ocean formation in icy bodies \cite{Postberg2011, Vance2021, zolotov2007, zolotov2012}. 

Despite this abundance, most interior models still assume \ce{NaCl} is excluded from high-pressure ice polymorphs and partitions entirely into the liquid phase — an assumption with direct consequences for predicted density profiles, compositional layering, and convective transport within icy mantles. Whether salts can instead be incorporated into the dense ice phases stable at several gigapascals — pressures typical of the deep interiors of large icy moons such as Ganymede and Titan \cite{Lger2004, SOTIN2007} — has remained an open question.

At low pressures ($<0.3$\,GPa), the answer seems clear: \ce{NaCl} is consistently rejected upon freezing \cite{Vrbka2005,Tsironi2020}, and the only known crystalline hydrate, hydrohalite (\ce{NaCl * 2 H2O}), is stable only near ambient pressure and low temperature \cite{Hamp2024, Mirwald2005}. 
Even here, though, the picture is more complex than a simple binary of `excluded or not'. Rapid freezing of \ce{NaCl} brine at ambient pressure can produce metastable and amorphous \ce{NaCl - H2O} phases beyond hydrohalite, with the resulting solid-phase assemblage recording the cooling history of the parent fluid \cite{Hamp2026}. 
At moderate pressures (< 2\,GPa) and low temperature (150–300\,K), hyperhydrated phases such as \ce{2NaCl * 17 H2O} (SC8.5) and \ce{NaCl * 13 H2O} (SC13) \cite{Journaux2023} show that \ce{NaCl} can be stoichiometrically incorporated into a water-ice framework \cite{Journaux2023}. SC13 even adopts an ice-VI-like hydrogen-bond network stabilised by ionic interactions \cite{Yamashita2023}. But these phases destabilise near the ice VI--ice VII transition ($\sim 2.2$\,GPa) \cite{Journaux2023}, leaving the fate of \ce{NaCl} in deeper ice VII mantles unknown. 

Above this transition, the response of ice VII to dissolved \ce{NaCl} turns out to depend strongly on both concentration and thermodynamic path \cite{Ludl2015, Ludl2017}. At low concentrations (1–5\,wt\%), ice VII forms a homogeneous phase with a slightly contracted lattice relative to the pure system \cite{Frank2006, Frank2008, Journaux2017}. This behaviour is consistent with ions acting as isolated point defects, locally distorting the hydrogen-bond network without altering the global structure. At elevated temperatures ($>500$\,K), this dilute regime breaks down via halite precipitation, suggesting limited long-term stability of dissolved salt under some conditions \cite{Frank2008}. Computational studies suggest small amounts of \ce{NaCl} can nonetheless be accommodated substitutionally within the ice VII lattice, extending the superionic regime and enabling ionic transport \cite{Hernandez2022}, while spectroscopy reveals shifts in the hydrogen-bond symmetrisation transition in its presence \cite{Bove2015}. None of this, however, tells us what happens at higher concentrations, or whether a qualitatively different, collective incorporation regime exists and which structure is formed.

\begin{figure}
    \centering
    \includegraphics[width=0.95\linewidth]{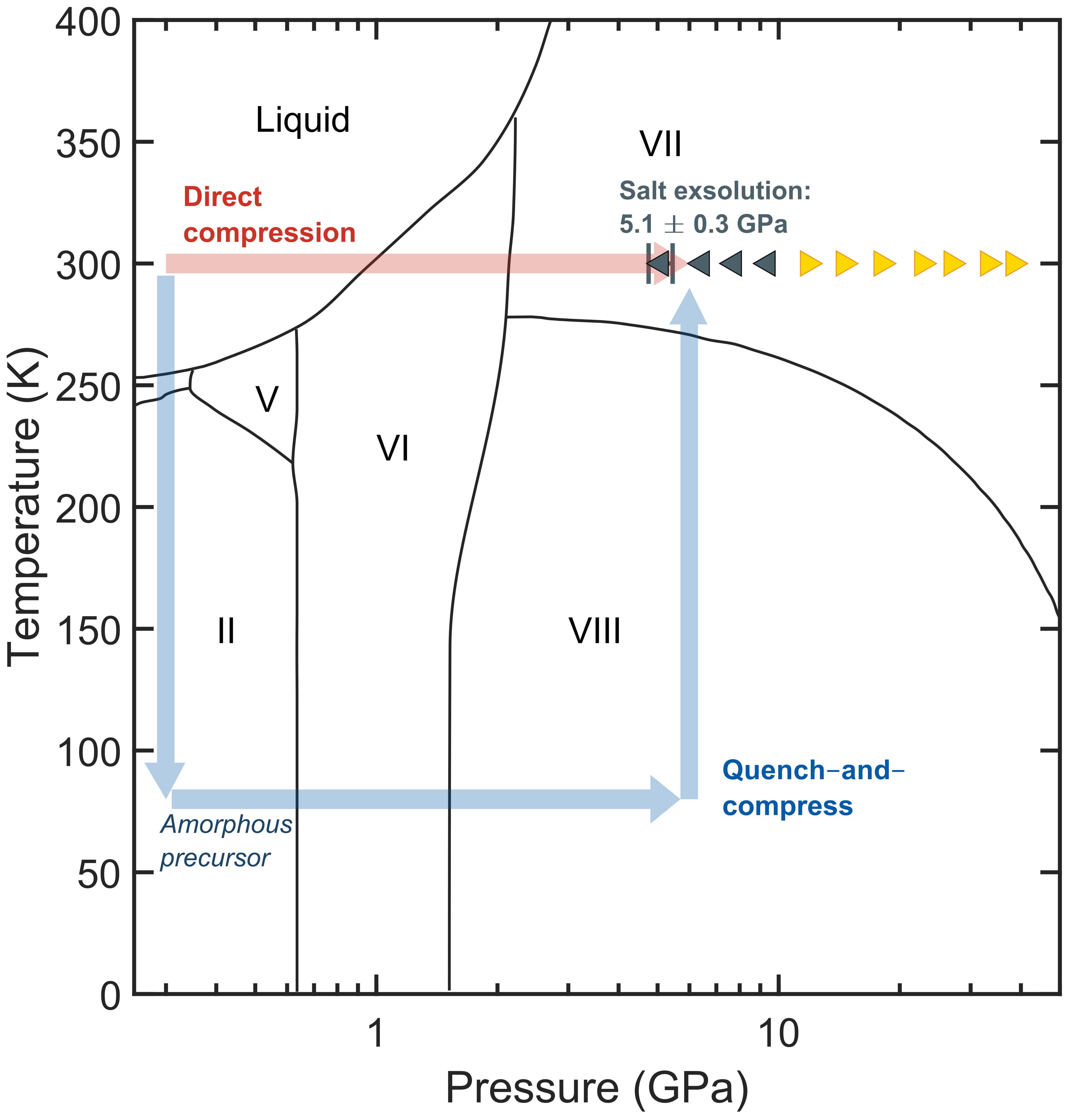}
    \caption{
    Phase diagram of pure \ce{H2O} with two different thermodynamic paths to reach the ice VII regime, indicated by the blue arrows and the red arrow. 
    Path No.1 (blue) follows the thermodynamic pathway prescribed in Ref.~\cite{Klotz2009}, called `temperature annealing' or `quench-and-compress' whilst path no.2 (red) is a direct room temperature compression, named `pressure annealing' or `direct compression'. 
    Selected XRD measurement points are shown for visual clarity, including the lowest and highest pressures investigated. The grey points represent XRD data taken upon decompression where we observed the sample desalting whilst the yellow points represent XRD data taken whilst measuring the Equation of State of the sample up to $\sim40$\,GPa. 
    XRD measurements were performed at the ID15b beamline at ESRF, Grenoble, France.}
    \label{Phase_diagram}
\end{figure}

A clue comes from the \ce{LiCl - H2O} system. In \ce{LiCl}-bearing ice VII, rapid quenching ($\sim10^2$\,Ks$^{-1}$) of concentrated solutions followed by compression above $\sim 4$\,GPa — a route we refer to hereafter as the `quench-and-compress' method — produces a distinct salty ice VII phase with an expanded lattice and suppressed proton ordering \cite{Klotz2009, Klotz2016}. This pathway relies on the formation of an amorphous precursor in which the ions locally achieve the coordination environment they will eventually need in the crystal \cite{Ludl2017}. Crystallisation can then proceed through local relaxation around these pre-configured environments, rather than requiring a collective reorganisation of the whole structure after the fact. For \ce{NaCl} in ice VII, both \ce{Na+} and \ce{Cl-} require  eightfold coordination, reached only at higher pressures ($\sim 5$\,GPa) \cite{Ludl2017, Ludl2015} than for \ce{LiCl}-bearing ice VII. Whether the same quench-and-compress pathway (Fig.~\ref{Phase_diagram}) can stabilise a true \ce{NaCl} substitutional solid solution in ice VII, and whether doing so drives a transition from defect-like incorporation to collective structural modification, has remained unresolved until now.

Here we demonstrate that a large amount of \ce{NaCl} can be extensively incorporated into ice VII as a substitutional solid solution, metastable above $5.1 \pm 0.3$\,GPa at 300\,K. Combining synchrotron X-ray diffraction, neutron scattering, and \textit{ab initio} calculations, we identify a previously unknown \ce{NaCl}-bearing ice VII phase with a unit cell $8.5$\,\% larger than pure ice VII and no detectable crystalline NaCl anywhere in the sample. Together with the observed lattice expansion and its reproduction by substitutional \textit{ab initio} models, this provides strong evidence for extensive ionic incorporation into the ice VII lattice. The results reveal a transition from dilute-defect behaviour to a collective substitutional regime, controlled by both concentration and the thermodynamic pathway.

This has a direct bearing on planetary interior models. If salt can be locked into ice VII rather than expelled at depth, the density and mechanical properties of salty high-pressure ice mantles — and with them, predictions of density stratification, convective dynamics, and long-term chemical transport in ocean worlds — need to be reconsidered. As we show below, this is particularly true near the base of the ice mantle, where salt may become trapped in a substitutional solid ice form rather than reaching the surface ocean.

% =================================================================
% ///////////////////////   METHODS   /////////////////////////////
% =================================================================

\section{Experimental Methods and Materials}
\subsection{Sample Preparation}
 
For X-Ray diffraction (XRD) experiments, a eutectic \ce{NaCl} solution ($25$\,wt\%, \ce{NaCl * 9.7 H_2O}) was prepared using \texttt{Milli-Q} water with $99.99$\,\% purity and \ce{NaCl} from \texttt{Sigma-Aldrich} with a purity $\geq 99.5$\,\%. The solution was both prepared and loaded into a membrane Diamond Anvil Cell (DAC) under a dry Nitrogen atmosphere.  
 
Type \textit{IIas} diamond anvils with culet diameters of $250$ and $500$\,\textmu m were used in our DACs. Rhenium-foil gaskets were used to contain the sample whilst ruby and gold spheres were used as pressure calibrants \cite{Mao1986, Chervin2001, Fei2007}. The equation of state of ice VII \cite{Klotz2017, Loubeyre1999, Lai2023} was also used as an additional pressure calibrant for the XRD experiments. 
The sample was formed by following the quench-and-compress thermodynamic path as shown in Fig.~\ref{Phase_diagram}, and described previously. The sample was compressed to $0.3$\,GPa before the entire DAC was plunged into a bath of liquid Nitrogen ($77$\,K), quenching the solution to form an amorphous precursor. At this cryogenic temperature, the sample was compressed to the target pressure, between $6 - 10$\,GPa before then being warmed up to room temperature. The pressure was carefully monitored upon warming to ensure that the sample remained in the target pressure range of above 6\,GPa. 
 
\subsection{X-Ray Diffraction}
 
Angle-dispersive X-ray powder diffraction patterns were acquired at both the ID15b and ID27 beamlines at the European Synchrotron Radiation Facility (ESRF, Grenoble, France) using a monochromatic X-ray beam ($\lambda$ = 0.4096\,\AA{} for ID15b, $\lambda$ = 0.3738\,\AA{} for ID27) and an \texttt{Eiger2 9M} \ce{CdTe} flat-panel detector. A typical acquisition time of 10\,s \cite{Garbarino2024, Poreba2023} was used.
The beam spot size for both instruments was $\sim \, 6\times 6$ \,\textmu m, and the DAC was rocked $5-10\,\degree$ during the acquisition.
The 2D diffraction patterns were treated in \texttt{Dioptas} \cite{Prescher2015} to mask the Bragg peaks of the diamond anvils, then integrated into powder-averaged one-dimensional patterns, and a smooth polynomial background was subtracted. Subsequent Le Bail fittings were performed using the \texttt{FullProf Suite} \cite{Rodrguez-Carvajal1993}. 
In the XRD measurements, the pressure was determined using the equation of state of gold \cite{Fei2007}, the equation of state of ice VII \cite{Klotz2017, Lai2023} or ruby fluorescence \cite{Mao1986, Chervin2001}%, and from the Raman spectra of the stressed diamond edge \cite{Akahama2006}. 
, depending on the experimental conditions. Where multiple calibrants were available, the resulting pressures were cross-checked to ensure consistency.

\subsection{Neutron Diffraction}
 %\textcolor{ForestGreen}{
 For the neutron diffraction experiment at the D20 beamline, Institut Laue-Langevin (ILL), the amorphous precursor was prepared using a high-pressure spray-injection technique as described by Ludl \textit{et al.} \cite{Ludl2017Quench}. A solution of $50$\,mm$^3$ of \ce{NaCl * 10.2 D2O} was sprayed into an evacuated chamber (order of 1\,mbar) and deposited onto a plate which was kept at $\sim80$\,K, enabling rapid vitrification and retention of the dissolved salt. Sample quality was checked using the XRD platform at the Institut de Min\'eralogie, de Physique des Mat\'eriaux et de Cosmochimie (IMPMC, Paris, France), with the sample under vacuum and at 77\,K. Once the amorphous precursor was confirmed, it was stored in liquid nitrogen so the \ce{NaCl} would not crystallise out of the amorphous. The sample had to be prepared in this way since the Paris-Edinburgh (PE) press used in high-pressure neutron diffraction requires several tens of mm$^3$ of sample. On site, the amorphous precursor was cryogenically loaded ($80$\,K) into a PE high-pressure clamp with cubic Boron Nitride (c\ce{BN}) anvils\cite{Klotz2013}, along with a piece of lead (Pb) acting as a pressure calibrant \cite{Strssle2014, Dewaele2019} It was then placed into the PE press and compressed to the target pressure ($6 - 10$\,GPa), all whilst at $80$\,K. The entire press and sample were then allowed to warm up to ambient temperature. This way, we followed the quench-and-compress thermodynamic pathway as shown in Fig.~\ref{Phase_diagram}. Measurements were done using a wavelength of $1.54$ or $2.41$\,\AA, depending on the desired resolution. Pressure was determined from the equation of state of \ce{Pb} \cite{Dewaele2019} and ice VII \cite{Klotz2017}. 
 Subsequent data refinement was carried out using the \texttt{FullProf Suite} \cite{Rodrguez-Carvajal1993}.
 %}
 
\subsection{Computational Details}
 
All first-principles calculations were performed using the Vienna Ab initio Software Package (VASP) \cite{Blchl1994, Kresse1996, Kresse1993, Kresse1999}, based on density functional theory (DFT), projector-augmented wave (PAW), and pseudopotentials. The generalised gradient approximation (GGA) with the Perdew-Burke-Ernzerhof (PBE) pseudopotentials was adopted to describe the exchange-correlation function \cite{Perdew1996}. 
 
The initial structure of ice VII was generated through \texttt{Genice2} \cite{Matsumoto2018} . Salty ice VII structures were constructed by randomly substituting water molecules with an equal number of \ce{Na+} and \ce{Cl-} ions. For $0$\,K static calculations, we set a plane-wave energy cutoff of $700$\,eV, and a $2\times 2 \times 2$ Monkhorst-Pack k-point grid mesh was adapted to sample the Brillouin zone. 
%\textcolor{ForestGreen}{
We evaluated equations of state (EoS) for four compositions: pure ice VII, and salty ice VII with \ce{NaCl} contents of $6$\,wt\% (1 \ce{Na+ - Cl-} ion pair replacing water molecules in the 54 molecule supercell), $19$\,wt\% and $35$\,wt\% (respectively 1 and 2 \ce{Na+ - Cl-} ion pairs replacing water molecules in the 16 molecule supercell).
%}
For the 19 and 35 wt\% composition, we randomly generated $20$ independent substitutional configurations. For the $6$\,wt\%, we neglected the hydrogen positions and identified seven distinct substitutional configurations. All configurations were fully optimised at each pressure with residual forces converging within $10-3$\,eV/\AA{}, and the EoS volumes were taken from the average of the optimised configurations.
 
The volumes of pure ice VII at $300$\,K were obtained from \textit{ab initio} molecular dynamics (AIMD). The AIMD initial structure has $128$ water molecules. We employed a timestep of $1$ femtosecond (fs), and the simulations lasted at least $10$ picoseconds (ps). At each time step, the total energy is converged to $0.1$\,meV. Simulations were carried out in the NVT ensemble with fixed composition (N), volume (V), and temperature (T). The temperature was controlled using the Nos\'e‐Hoover thermostat \cite{Hoover1985, Nos1984}. 
%\textcolor{ForestGreen}{
Additional AIMD simulations were performed for \ce{NaCl}-bearing ice VII containing 25\,wt\% and 35\,wt\% \ce{NaCl}, to evaluate the effect of salt incorporation on the thermal expansion. The salty ice structures were constructed from a 128-\ce{H2O} supercell by randomly replacing water molecules with equal numbers of \ce{Na+} and \ce{Cl-} ions (11 Na and 11 Cl for 25\,wt\% and 16 Na and 16 Cl for 35\,wt\%).
%}
A single substitutional configuration was used for each composition in the AIMD calculations.  The average thermal expansion coefficients $\alpha = \frac{ln(V_{300\,K}/V_{0\,K})}{300}$ at pressures were estimated based on the EoS from AIMD ($300$\,K) and static calculation ($0$\,K). From simulations, we saw that salt incorporation had a negligible effect on $\alpha$ (Supplementary Fig.\textcolor{blue}{S4}). 
%\textcolor{ForestGreen}{
Therefore, 
%} 
the $0$\,K volumes of salty ices were scaled to $300$\,K based on the 
%\textcolor{ForestGreen}{
pressure-dependent 
%}
$\alpha$ of pure ice VII. The resulting estimated $300$\,K volumes and $\alpha$ are reported in Table~\ref{Tab Xi Volumes and alpha}.
 
\begin{table}[h]
\centering
\renewcommand{\arraystretch}{1.5}
\begin{tabular}{c c c c c}
\hline
\multirow{2}{4em}{Pressure (GPa)} & \hspace{1cm} & Volume (\AA$^3$) & \hspace{1cm} & \multirow{2}{4em}{\shortstack{$\alpha$ \\ $(10^{-5}\,\mathrm{K}^{-1})$}} \\
\cline{2-4}
& Pure ice VII & $19$\,wt\% & $35$\,wt\% & \\
\hline
$5$ & $35.32$ & $38.26(24)$ & $40.21(17)$ & $15.0$ \\
$10$ & $31.72$ & $33.30(75)$ & $34.99(13)$ & $9.5$ \\
$20$ & $27.80$ & $28.89(43)$ & $30.00(74)$ & $5.1$ \\
$30$ & $25.47$ & $26.37(29)$ & $27.31(57)$ & $3.3$ \\
$40$ & $23.83$ & $24.70(25)$ & $25.53(44)$ & $2.2$ \\
% (235) -> (24), (168) -> (17), (134) -> (13)
\hline
\end{tabular}
\caption{Estimated $300$\,K volumes of pure and \ce{NaCl}-ice VII structures (reported uncertainties reflect only configurational variation in the 0\,K volumes, not including the additional uncertainty associated with the $\alpha$.), as well as the average thermal expansion coefficient $\alpha=\frac{ln(V_{300\,K}/V_{0\,K})}{300}$.}
\label{Tab Xi Volumes and alpha}
\end{table}

% ==================================================
% ///////////////     RESULTS     //////////////////
% ==================================================

\section{Results}
\subsection{Discovery of a \ce{NaCl}-bearing ice VII phase}

\begin{figure*}
    \centering
    \includegraphics[width=0.95\textwidth]{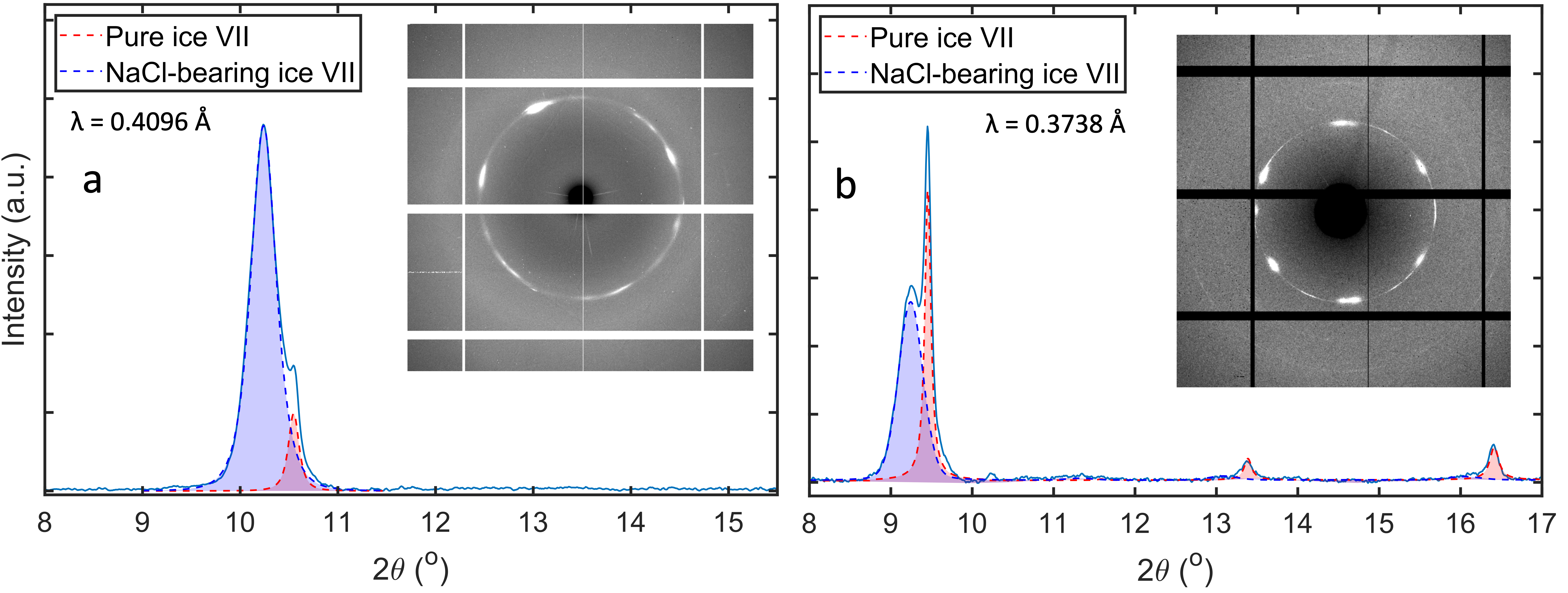}
    \caption{(a) Representative X-ray diffractograms of two different \ce{NaCl}-bearing ice VII samples at $300$\,K. (a) Sample of $26.9$\,wt\% \ce{NaCl} at $9.62$\,GPa in a DAC with $250$ micron culets, measured at ID15b, ESRF with a wavelength $\lambda = 0.4096$\,\AA. (b) A second sample of $33.1$\,wt\% \ce{NaCl} at $7.67$\,GPa in a DAC with $500$ micron culet diamonds at ID27, ESRF with a wavelength $\lambda = 0.3738$\,\AA. In both panels, the blue shaded Bragg peaks are for the \ce{NaCl}-bearing ice VII phase whilst the red shaded peaks indicate pure ice VII. 
    \textit{Insets:} The recorded two-dimensional diffraction plate images of the respective measurements. There is a complete powder ring from the pure ice VII, and a partially ordered powder ring from the \ce{NaCl}-bearing ice VII present in both images.}
    \label{XRD Cell Expansion}
\end{figure*}

%\textcolor{ForestGreen}{
Following the quench-and-compress path (Fig. ~\ref{Phase_diagram}) with the $25$\,wt\% \ce{NaCl} solution and subsequent warming to room temperature, synchrotron XRD at the ID27 beamline (ESRF) revealed a previously unobserved, expanded ice VII phase coexisting with pure ice VII at $7.67$\,GPa, as determined by temperature-corrected ruby fluorescence \cite{Shen2020, Chervin2001}. Both phases adopt the cubic $Pn\overline{3}m$ space group but have distinct unit-cell parameters (Fig.~\ref{XRD Cell Expansion}b).
%} 
One gives complete Debye–Scherrer rings, consistent with pure ice VII whilst the other gives partially ordered rings, suggesting uniaxial strain within the crystal. Critically, mapping the full sample volume, no crystalline \ce{NaCl}, nor any \ce{NaCl}-hydrate, was detected. Together with the expanded ice VII lattice and the agreement with the substitutional calculations, this observation indicates extensive incorporation of Na$^+$ and Cl$^-$ into the ice VII structure. 

%\textcolor{ForestGreen}{
A Le Bail fitting scheme was used to determine the unit-cell and peak-profile parameters of the two ice VII phases without imposing a detailed atomic model. These refinements show that the \ce{NaCl}-bearing ice VII phase has a unit cell $\sim8.5\,\%$ larger than the pure, coexisting ice VII. The refinement further also shows that the diffraction peaks of the \ce{NaCl}-bearing ice VII phase are $4.5$ times broader than those of the pure ice VII, indicating substantially more heterogeneous microstructure in the expanded phase, discussed further in Section IV.B.
%}

A separate constrained Rietveld refinement of this ID27 diffraction data was used for the limited purpose of estimating the oxygen isotropic displacement parameter. The refined value increased from \(B_{\mathrm{iso}}=1.05(2)\,\text{Å}^2\) in the coexisting pure ice VII phase to \(11.8(9)\,\text{Å}^2\) in the \ce{NaCl}-bearing ice VII phase. Rietveld refinement was not used to determine site occupancies or a detailed atomic arrangement. Hydrogen positions and their displacement parameters were not refined because of the limited sensitivity of X-ray diffraction to hydrogen. Given the substantial substitutional disorder and the correlations between displacement, occupancy and profile parameters, the increased oxygen \(B_{\mathrm{iso}}\) should therefore be regarded as an effective indicator of enhanced structural disorder rather than as a quantitative measure of atomic thermal motion.

The expanded lattice, together with the absence of crystalline \ce{NaCl} or any \ce{NaCl}-hydrate, indicates that NaCl is incorporated within the ice VII structure itself rather than forming a separate phase.

%\textcolor{ForestGreen}{
It is important to note though that the relative proportions of pure and expanded ice VII differ between samples prepared using different DAC geometries. DACs with diamonds with a $500$\,\textmu m culets and another with culets of $250$\,\textmu m were both used. Both samples were prepared with the same $25$\,wt\% \ce{NaCl} starting solution. The sample within the DAC with $500$\,\textmu m culets contained a greater proportion of pure ice VII than the sample prepared with the DAC with $250$\,\textmu m (See Fig.~\ref{XRD Cell Expansion}).
Analysis of the relative XRD intensities, corrected for the scattering-power factors of the two phases, indicates that the local \ce{NaCl}-bearing ice VII concentrations are approximately $26.9$\,wt\% (\ce{NaCl * 8.82 H2O}, Fig.~\ref{XRD Cell Expansion}a) for the DAC with 250\,\textmu m diamonds and $33.1$\,wt\% (\ce{NaCl * 6.56 H2O}, Fig.~\ref{XRD Cell Expansion}b) in the DAC with 500\,\textmu m culets. We hypothesise that the difference arises from the difference in the cooling rate of the two different DAC geometries, controlling how much \ce{NaCl} is retained in the amorphous precursor before crystallisation, although further tests would be required to confirm this.
%}

%\textcolor{ForestGreen}{
Additional experimental support of the \ce{NaCl}-bearing ice VII was obtained from neutron diffraction measurements at the D20 beamline at the Institut Laue-Langevin (ILL), Grenoble, France. The amorphous precursor was produced and cryoloaded as described in the methods section II.C. Measurements were taken whilst the amorphous was warmed to form crystalline ice VII (Supplementary Fig.\textcolor{blue}{S1}). 
The resulting neutron diffraction patterns collected at room temperature and shown in Fig.\textcolor{blue}{S2}, again reveal two cubic lattices and no crystalline \ce{NaCl}. However, in this case, the expanded ice VII unit cell volume is greater than that of pure ice VII by only $\sim3$\,\%. This expansion is significantly smaller than the previously mentioned XRD experiments. 
During the spray-quenching process small, cubic crystalline inclusions were observed to have formed, but not collected, as their identity was uncertain at the time. Since a reduced lattice expansion was measured with neutrons, it is plausible that these crystallites corresponded to crystalline \ce{NaCl} which had separated during the spray-quenching process. If salt segregation during quenching had occurred, the effective \ce{NaCl} concentration incorporated into the amorphous precursor, and thus into the ice VII lattice, would have been lower than the original solution concentration, resulting in a smaller expansion and placing the sample into the dilute regime.
%}

\subsection{Local structure of the salty lattice}

\begin{figure*}
    \centering
    \includegraphics[width=0.95\textwidth]{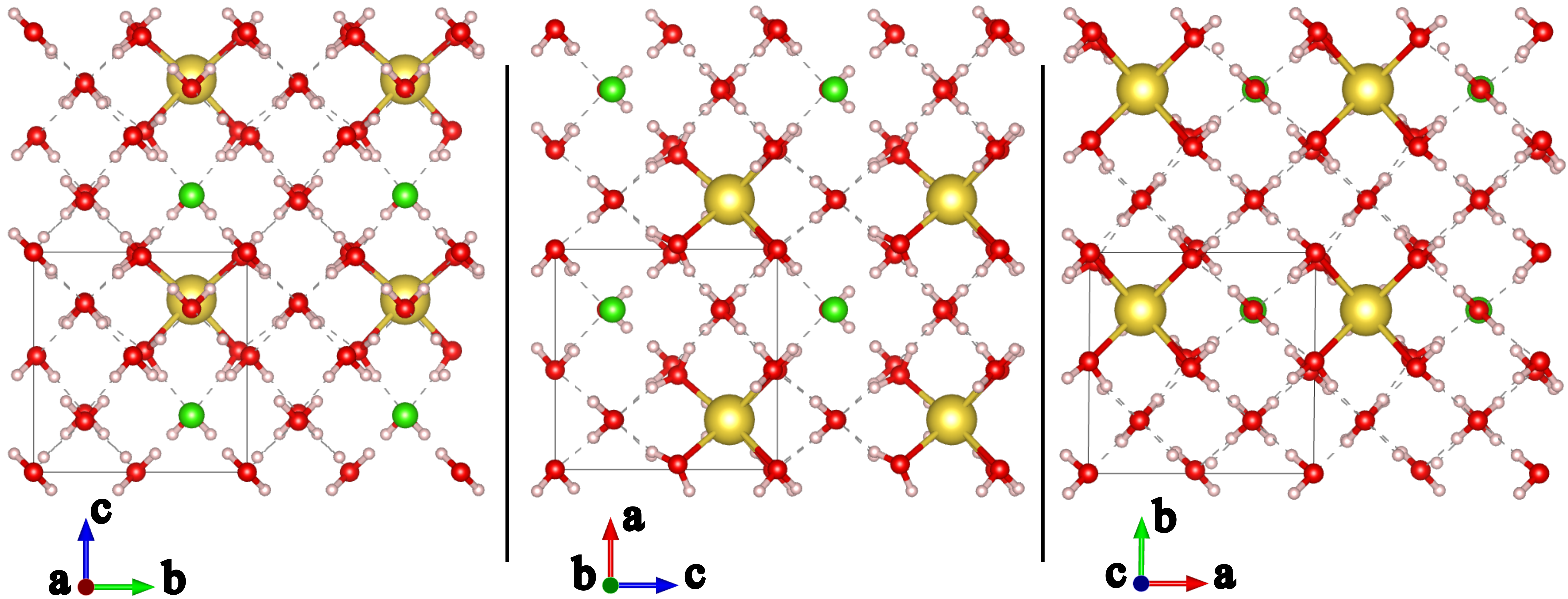}
    \caption{Candidate structure of \ce{NaCl}-bearing ice VII with $19$\,wt\% \ce{NaCl} from \textit{ab initio} calculations. The red spheres represent locations of the oxygen ions, grey for hydrogen ions, yellow for sodium ions and green for chlorine ions. The three panels correspond to looking down the a-, b- and c-axes respectively. The sodium and chlorine ions are each substituting a water molecule within the ice VII framework.}
    \label{Crystal Structure}
\end{figure*}

Ab initio calculations (VASP) were used to model the expanded ice VII lattice at two NaCl concentrations, $19$\,wt\% and $35$\,wt\%, substituting water molecules for \ce{Na+} and \ce{Cl-} pairs within the ice VII supercell prior to relaxation. The  $19$\,wt\% composition is used as a representative case for discussing local structure.
Configurations in which \ce{Na-Cl} distances approached $2.7$\,\AA{} — the \ce{Na-Cl} distance in crystalline \ce{NaCl} at these pressures \cite{Dewaele2019} — were excluded. Their simulated diffraction patterns show strong low-angle Bragg reflections that are absent from the experimental data, strongly disfavouring configurations containing substantial \ce{Na - Cl} pairing, resembling embryonic \ce{NaCl}.
The remaining, non-ion-paired configurations preserve the ice VII topology, despite Na$^+$ and Cl$^-$ each substituting for a water molecule (Fig.~\ref{Crystal Structure}). These structures show only weak low-angle reflections, which diminish further with increasing disorder, consistent with the broadened, largely featureless diffraction observed experimentally. The local hydrogen-bond network reorganises around each ion. The hydrogens of the water molecules orientate toward neighbouring \ce{Cl-}, while oxygen atoms orientate towards \ce{Na+}. The remaining water network preserves the Bernal–Fowler ice rules away from the substituted sites. The model is consistent with the experimental diffraction observations and provides a structural mechanism by which the ice framework accommodates the incorporated ions.

\subsection{Pressure stability and the desalting transition}

\begin{figure}[!b]
\centering
\includegraphics[width=0.95\columnwidth]{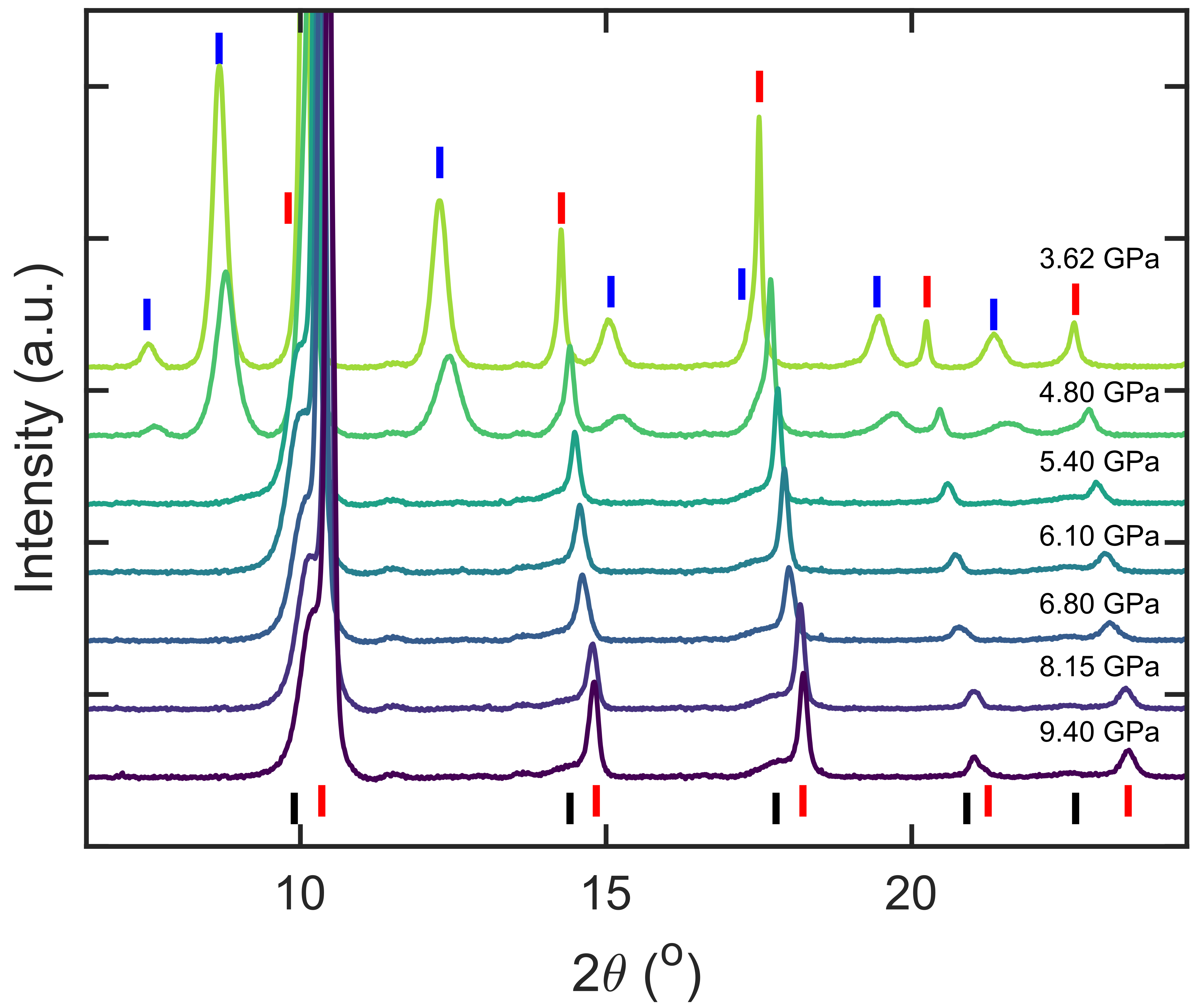}
\caption{XRD reveals that pure crystalline \ce{NaCl} is expelled from the \ce{NaCl}-bearing ice VII structure between $5.40$\,GPa and $4.80$\,GPa. Red tick marks indicate the peaks belonging to pure ice VII, black tick marks are for the \ce{NaCl}-bearing ice VII and blue tick marks indicate peaks belonging to the pure crystalline \ce{NaCl}-B1 phase.}
\label{Desalt}
\end{figure}

To determine the pressure range over which the substitutional phase is stable, a third, independently prepared \ce{NaCl}-bearing ice VII sample was prepared following the quench-and-compress path in a DAC with $500$\,\textmu m culets, before being warmed up at $9.40$\,GPa. Subsequently, the sample was decompressed with an XRD pattern collected at each step (Fig.~\ref{Desalt}). The salty phase persists unchanged down to $5.40$\,GPa. At $4.80$\,GPa, diffraction peaks corresponding to crystalline \ce{NaCl} appear at the same time as the disappearance of the expanded ice VII structure and the growth of the pure ice VII peaks, marking the onset of phase separation into pure ice VII and exsolved salt, a process referred to hereafter as `desalting'. These two observations bracket the desalting pressure at $5.10 \pm 0.30$\,GPa.
The transition is irreversible upon recompression: increasing the pressure above $5.10$\,GPa again does not redissolve the exsolved \ce{NaCl}, which remains crystalline. \ce{NaCl}-bearing ice VII is therefore a metastable, rather than thermodynamically reversible, solid solution. This asymmetry is relevant to salt transport in planetary ice mantles, as discussed further in Section IV.C. 

\subsection{Equation of State and Density of \ce{NaCl}-bearing ice VII}
 
\begin{figure}
\centering
\includegraphics[width=0.95\columnwidth]{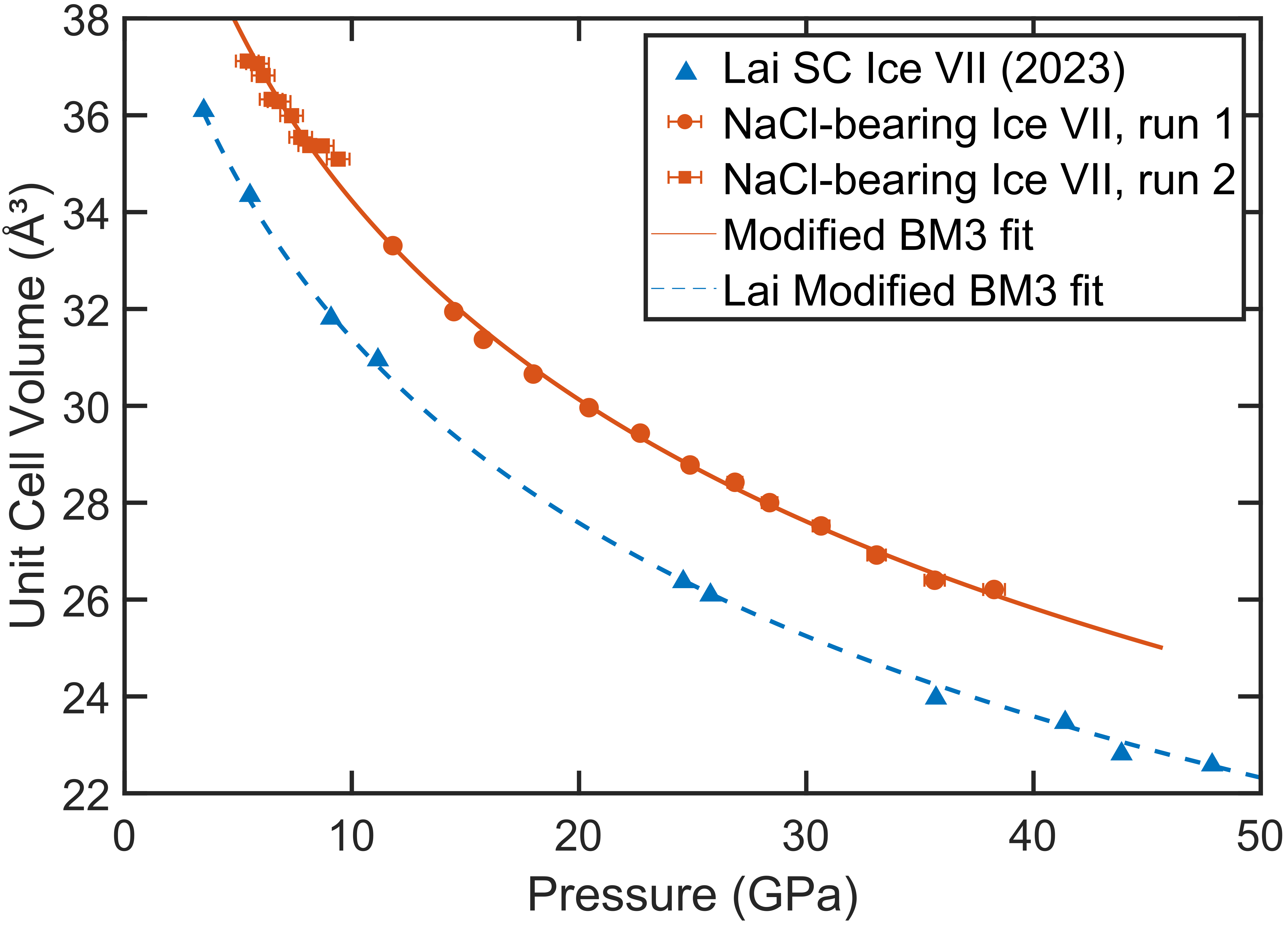}
\caption{Isothermal equation of state of \ce{NaCl}-bearing ice VII measured at the ID15b beamline at ESRF, Grenoble. The $26.9$\,wt\% \ce{NaCl}-bearing ice VII data (red) are shown with the fitted modified third-order Birch--Murnaghan equation of state (solid red line), using a reference pressure of $P_R = (5.1 \pm 0.3)$\,GPa. The fitted parameters are $V_R = (37.81 \pm 1.70)$\,{\AA$^3$}, $K_R = (40.85 \pm 0.21)$\,GPa, and $K_R' = (4.30 \pm 0.24)$. 
Blue triangles show XRD literature data for single-crystal pure ice VII from Lai \textit{et al.}~\cite{Lai2023}, together with the corresponding modified third-order Birch--Murnaghan equation of state fitted here using the same reference pressure, $P_R = (5.1 \pm 0.3)$\,GPa.}
\label{EoS}
\end{figure}

The $250$\,\textmu m culet sample introduced in Section III.A and shown in Fig.~\ref{Crystal Structure}a was subsequently used to measure isothermal equation of state of the $26.9$\,wt\% \ce{NaCl}-bearing ice VII between $7.3$ and $38.3$\,GPa (Fig.~\ref{EoS}). The coexisting pure ice VII phase as an internal pressure calibrant \cite{Klotz2017, Lai2023}. A modified third-order Birch-–Murnaghan EoS \cite{Lai2023} (Eq.~\Ref{mBM3}), fitted with a reference pressure of $P_R = 5.10$\,GPa, gives $V_R = (37.81 \pm 1.70)$\,{\AA$^3$}, $K_R = (40.85 \pm 0.21)$\,GPa and $K_R' = (4.30 \pm 0.24)$ with a reduced $\chi$-squared value, $\chi^2_{red.} = 1.35$, indicating good agreement with the experimental p–V data.

\begin{multline}\label{mBM3}
P - P_R = \frac{3K_R}{2}\left[\left(\frac{V_R}{V}\right)^{\frac{7}{3}} - \left(\frac{V_R}{V}\right)^{\frac{5}{3}}\right] \cdot \\ 
\left[1 + \frac{3}{4}(K_R'-4)\left(\left(\frac{V_R}{V}\right)^{\frac{2}{3}}\right) - 1\right]
\end{multline}

\begin{table*}[t]
\centering
\renewcommand{\arraystretch}{1.5}
\begin{tabular}{c c c c c c c}
\hline
Sample & $P-T$ Range & $K_R$ (GPa) & $K_R'$ & $V_R$ (\AA$^3)$ & Equation of State & Reference \\
\hline
Powder \ce{NaCl}-Ice VII & $5.4-38.3$\,GPa, RT & $40.85 (0.21)$ & $4.30(0.24)$ & $37.81 (1.70)$ & mBM3 & This study \\
SC Ice VII & $5.5-41.4$\,GPa, RT & $38.3$ & $4.23$ & $34.87$ & mBM3 & Lai \textit{et al.}, 2023 \cite{Lai2023} \\
SC Ice VII & $5.4-39.4$\,GPa, RT & $35.0$ & $4.44$ & $34.85$ & mBM3 & Loubeyre \textit{et al.}, 1999 \cite{Loubeyre1999} \\
\hline
\end{tabular}
\caption{Comparison of the fitted modified Birch--Murnaghan Equation of State parameters for single-crystal (SC) pure ice VII and \ce{NaCl}-bearing ice VII. The Loubeyre and Lai data were re-fitted with the same mBM3 with a reference pressure of $P_R = 5.1 \pm 0.3$\,GPa, so that the comparison is consistent.}
\label{Tab EoS Parameters}
\end{table*}

Comparison with pure ice VII fitted using the same reference pressure (Table~\ref{Tab EoS Parameters}) shows two systematic differences. The bulk modulus of the salty phase is modestly higher than the literature values obtained using the same EoS formulation, indicating a stiffening relative to pure ice VII. The reference volume is $8.5$\,\% larger, consistent with the lattice expansion established from the diffraction data. NaCl incorporation therefore expands both the lattice and increases its stiffness. The adherence of the sample to Vegard's law \cite{Denton1991} provides further support that the phase constitutes a true substitutional solid solution rather than a mechanical inclusion of salt. 

\begin{figure}
\centering
\includegraphics[width=0.95\columnwidth]{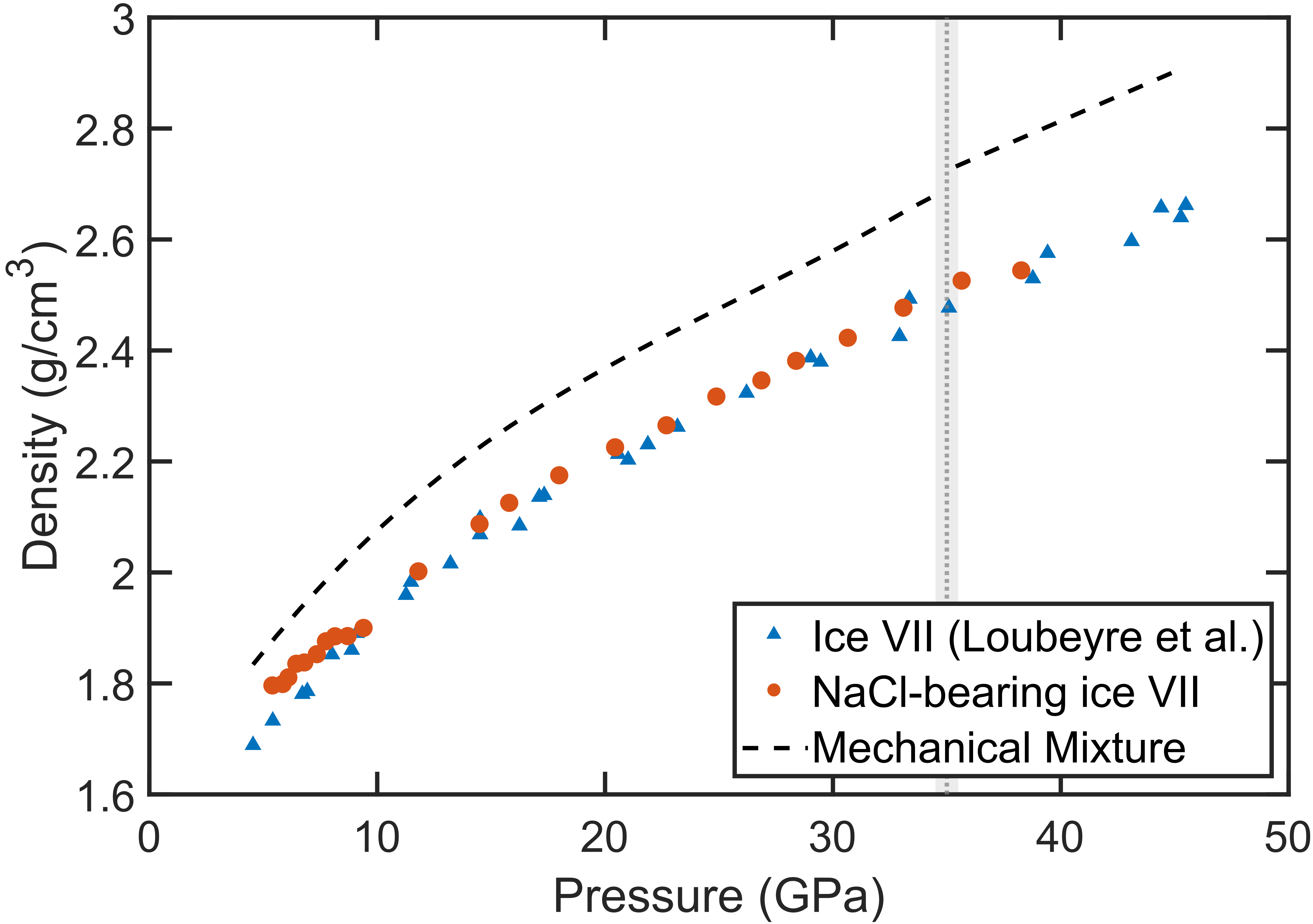}
\caption{The density of \ce{NaCl}-bearing ice VII as a function of pressure. The red points indicate the measured density of the $26.9$\,wt\% \ce{NaCl}-bearing ice VII, the blue triangles indicate literature data from Loubeyre \cite{Loubeyre1999} whilst the dashed black line indicates the calculated density for the mechanical mixture of ice VII and \ce{NaCl} with the same stoichiometry ($26.9$\,wt\%, \ce{NaCl * 8.82 H2O}). The vertical grey dotted line indicates the \ce{NaCl}-B1 to B2 phase transition at $35$\,GPa \cite{Dewaele2019}.}
\label{Density}
\end{figure}

The additional mass of the incorporated ions, along with the accompanying lattice expansion, act in opposite directions on density, motivating a direct comparison with pure ice VII. Converting the measured unit-cell parameters to density and comparing against both pure ice VII and a mechanical mixture of ice VII and \ce{NaCl} at equivalent bulk composition (\ce{NaCl * 8.82 H2O}, $26.9$\,wt\%), Fig.~\ref{Density} shows that the density of \ce{NaCl}-bearing ice VII closely follows that of pure ice VII across the full pressure range, despite the incorporation of $26.9$\,wt\% of \ce{NaCl}. The lattice expansion largely offsets the additional ionic mass. The mechanical mixture, by contrast, remains consistently denser than both phases and exhibits a distinct density increase at $35$\,GPa, corresponding to the bcc-B1 to fcc-B2 transition of crystalline \ce{NaCl}. \ce{NaCl}-bearing ice VII shows no equivalent discontinuity, because the \ce{Na+} and \ce{Cl-}  ions remain incorporated within the disordered ice VII framework rather than forming a separate crystalline phase.

% ================================================================
% \\\\\\\\\\\\\\\\\\\\\\\\\\\\\\\\\\\\\\\\\\\\\\\\\\\\\\\\\\\\\\\\
% ================================================================
 
\section{Discussion}

\subsection{Pathway- and concentration-dependent incorporation}

The key distinction between the \ce{NaCl}-bearing ice VII identified here and previously reported dilute salty ice VII lies not only in salt concentration, but also in the pathway by which crystallisation occurs. Direct compression of aqueous \ce{NaCl} solutions (`Direct compression' pathway, Fig.~\ref{Phase_diagram}), as in our XRD experiments (Fig. \textcolor{blue}{S3}) and previous work \cite{Frank2006}, produces ice VII without the pronounced lattice expansion obtained through the quench-and-compress route. A possible explanation is that crystallisation during direct compression precedes the development of the local ionic coordination favourable for extensive substitution. Rapid quenching instead traps the ions within an amorphous \ce{H2O} matrix, which can be compressed before crystallisation. At pressures above approximately $5$\,GPa this pathway favours the development of eightfold ionic coordination \cite{Ludl2017}, allowing the ice VII network to crystallise around preconfigured ionic environments during subsequent warming.

This interpretation accounts for the strong pathway dependence and the failure of the salt-bearing phase to reform after desalting. Once \ce{NaCl} exsolves and crystallises as halite upon decompression, simply increasing the pressure does not restore the dispersed ionic environments present in the compressed amorphous precursor. The observed irreversibility therefore highlights a substantial kinetic barrier to reincorporation. The metastable retention of the expanded phase above $(5.1 \pm 0.3)$\,GPa at 300\,K should consequently be distinguished from an equilibrium solubility boundary. 
%\textcolor{ForestGreen}{
We stress that the desalting pressure of $5.1 \pm 0.3$\,GPa is the experimentally determined kinetic desalting threshold at 300\,K, rather than a universal equilibrium boundary. Its location under planetary geotherms remains to be established.
%} 

\begin{figure}
\centering
\includegraphics[width=0.95\columnwidth]{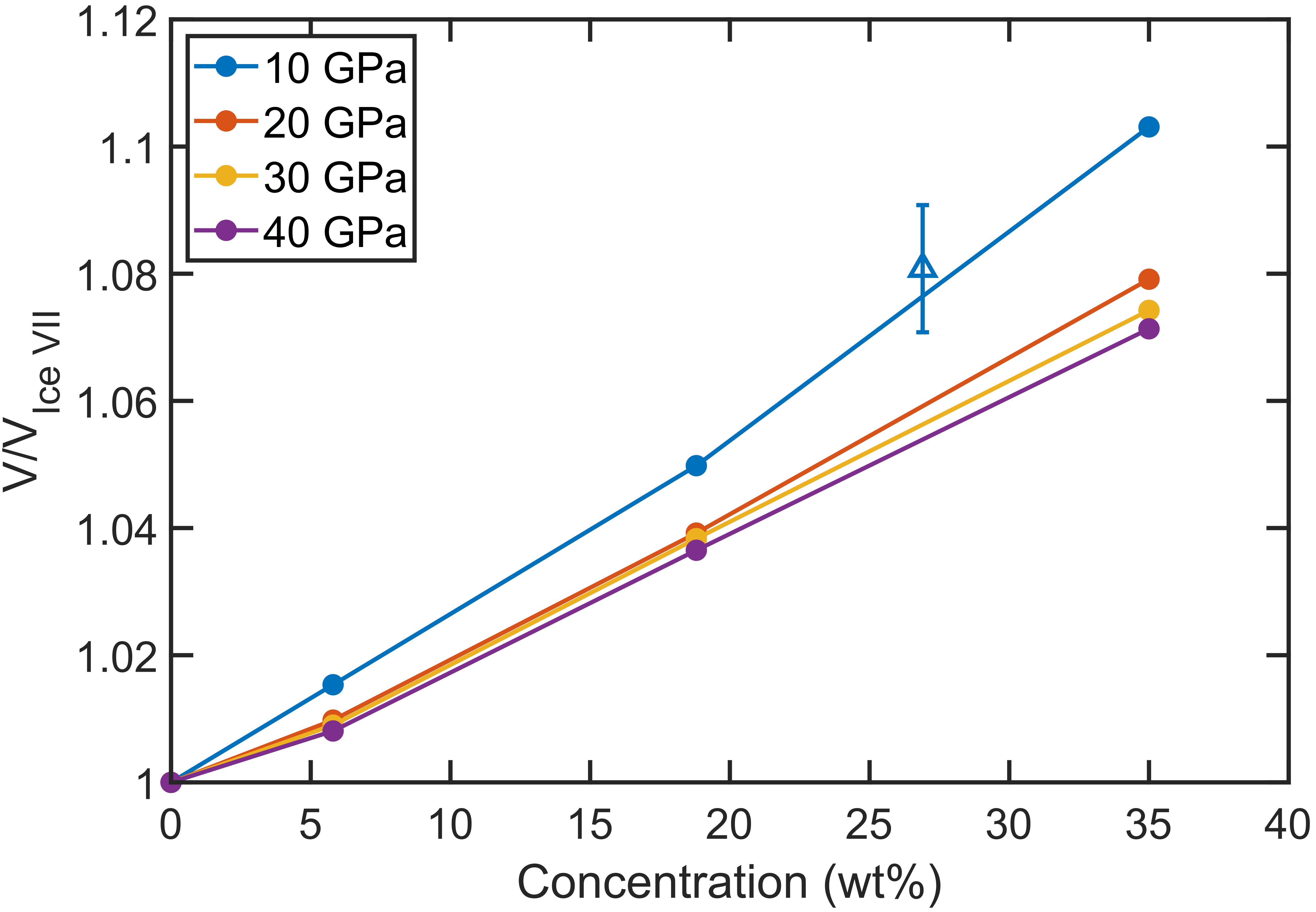}
\caption{Concentration dependence of lattice expansion in \ce{NaCl}-bearing ice VII at $10$\,GPa. Static DFT calculations were used to calculate the volume of \ce{NaCl}-bearing ice VII with $6$\,wt\%, $19$\,wt\% and $35$\,wt\% of \ce{NaCl} (filled circles), corrected for $300$\,K, and experimental data with $26.9$\,wt\% of \ce{NaCl} is also displayed (open triangle) measured at $9.62$\,GPa.}
\label{Conc}
\end{figure}

Salt concentration provides a second control on the structural response. Fig.~\ref{Conc} compares the calculated lattice expansion at approximately $6$, $19$ and $35$\,wt\% \ce{NaCl} with the experimental result at approximately $26.9$\,wt\%, showing an increasing expansion with salt content. At low concentration, individual ions can be viewed as isolated defects: they reorientate neighbouring water molecules and perturb local \ce{O-O} distances, while the surrounding hydrogen-bond network accommodates these distortions. This picture is consistent with the small lattice changes observed in dilute systems, including our neutron diffraction experiment (Fig.~\textcolor{blue}{S2}). Related local ionic distortions have also been reported in the ice-VI-like SC13 framework \cite{Yamashita2023}. At higher concentrations, the distorted environments surrounding neighbouring ions are expected to overlap, producing a collective perturbation of the water network rather than independent local relaxations. Analogous effects have been discussed for concentrated ionic solutions \cite{Winkel2011, Ruiz2014}.

The pronounced lattice expansion at high \ce{NaCl} content may thus reflect a crossover from isolated local distortions to a more collective response of the hydrogen-bond network, while preserving the average ice VII topology. A systematic experimental study as a function of salt concentration is needed to test this interpretation and establish how the structural response evolves between the dilute and concentrated limits. These results nevertheless highlight the importance of both composition and crystallisation history in determining how high-pressure ice accommodates salt.

\subsection{Structural disorder and physical properties}

Ionic incorporation preserves the average ice VII topology while modifying the local arrangement of water molecules. In the calculated structures (Fig. 3), neighbouring molecules reorientate in response to the ionic environment, with oxygen atoms directed preferentially toward \ce{Na+} and protons towards \ce{Cl-}. The spatial distribution of substitutional ions therefore introduces local orientational correlations and heterogeneous distortions within an otherwise cubic framework.

%\textcolor{ForestGreen}{
This microscopic picture is consistent with the diffraction observations. The substantial peak broadening and increased effective oxygen \(B_{\mathrm{iso}}\) provide complementary indications of structural disorder.
%} 
Peak broadening reflects heterogeneous strain and potentially a reduction in coherent-domain size with particles having strong inhomogeneities in the \ce{Na+}/\ce{Cl-} concentrations. Additionally, the enhanced displacement parameter indicates larger deviations from the average crystallographic positions. Together with the calculated local rearrangements, they support a strongly disordered salt-bearing lattice. Similar structural perturbations have been observed in LiCl-bearing ice VII \cite{Klotz2009, Klotz2016, Bove2015} and predicted for ionically doped proton-disordered ice \cite{Bronstein2016}.

Despite this pronounced local disorder and the $8.5$\,\%  increase in unit-cell volume, the bulk compressional response remains close to that of pure ice VII. The fitted bulk modulus of the \ce{NaCl}-bearing phase is modestly larger over the same pressure range (Tab. \ref{Tab EoS Parameters}), indicating a small reduction in compressibility. Ionic substitution therefore expands and locally distorts the lattice without making the resulting framework mechanically softer.

The most consequential physical effect is the balance between the additional ionic mass and the accompanying lattice expansion. For the composition investigated here, the increase in unit-cell volume largely compensates for the replacement of water molecules by the heavier \ce{Na+} and \ce{Cl-} ions. The density of \ce{NaCl}-bearing ice VII consequently remains close to that of pure ice VII throughout the investigated pressure range, whereas an equivalent mechanical mixture of ice VII and crystalline \ce{NaCl} is substantially denser (Fig. \ref{Density}). The mechanical mixture also displays the density discontinuity associated with the B1-–B2 transition of crystalline\ce{NaCl} near 35\,GPa, which is absent when the ions remain incorporated within the ice VII framework.

Thus, the same bulk salt content produces markedly different physical properties depending on whether \ce{NaCl} is incorporated within the ice lattice or segregated as crystalline halite.

\subsection{Implications for planetary interior dynamics and salt transport}

\begin{figure*}
    \centering
    \includegraphics[width=0.95\textwidth]{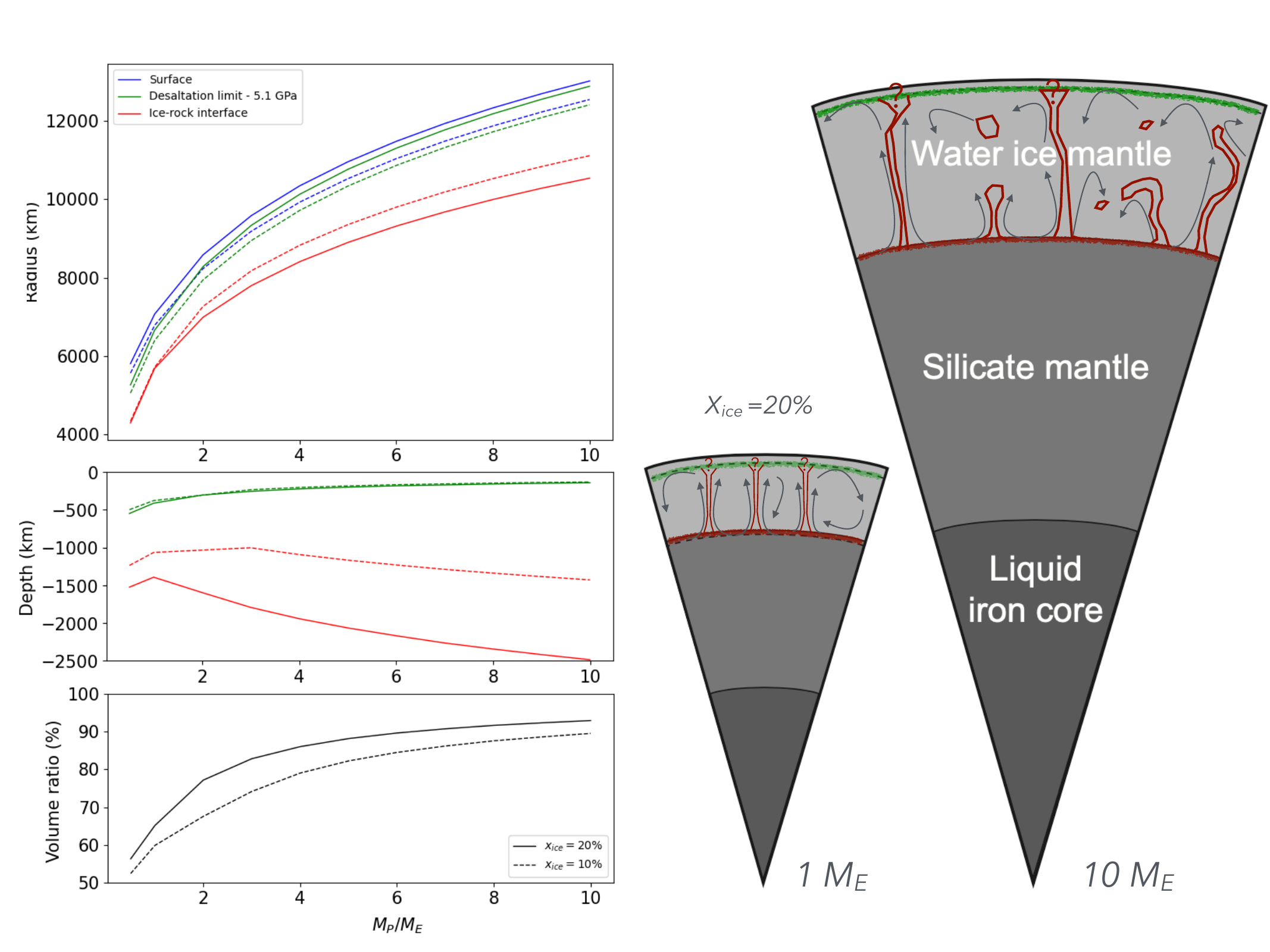}
    \caption{Surface radius (blue), radius/depth of the desalting transition at 5.1\,GPa (green), and the ice-rock interface (red) as a function of the planetary mass relative to that of the Earth, for differentiated icy planets having a total mass fraction of ice, $x_{ice}$, of 10\,\% (dashed line)  and 20\,\% (solid line). The internal structure profiles were calculated using the interior model of \cite{SOTIN2007}, with internal parameters similar to those in \cite{Tobie2019}. The bottom plot displays the volume ratio between the icy mantle above 5.1\,GPa in which \ce{NaCl}-bearing ice VII is metastably retained at $300$\,K, and the total volume of the icy mantle. Two examples of internal structures are shown on the right, for an ice mass fraction of 20\,\% and total masses of 1 and 10 Earth masses. The internal structures are divided into three main layers: a liquid iron core, a silicate mantle and an icy mantle. The green dashed line indicates the location of the desalting transition. Arrows and red contour lines illustrate the possible convective motions and the potential upward advection of \ce{NaCl}-bearing ice VII, if present within the deeper icy mantle, towards the desalting transition.} 
    \label{int_struct_salt}
\end{figure*}

The contrast between incorporated and exsolved salt has direct implications for transport through high-pressure ice mantles. Hernandez et al. \cite{Hernandez2022} predicted \ce{NaCl}-bearing ice VII to be denser than pure ice VII and comparable in density to the corresponding assemblage of ice VII and crystalline salt. Even under those conditions, their simulations showed that \ce{NaCl}-bearing ice VII could be efficiently entrained by mantle convection. Our results strengthen this conclusion by showing that the density of the substitutional phase is instead close to that of pure ice VII (Fig. \ref{Density}), implying only a limited compositional buoyancy contrast during transport through the deep mantle.

This density behaviour has two main consequences for salt transport. First, \ce{NaCl}-bearing ice VII formed or retained in the deep mantle could be transported upwards by solid-state convection without a major density penalty. Second, once the upwelling material crosses the \(5.1 \pm 0.3\) GPa desalting boundary, \ce{NaCl} exsolves from the ice VII lattice and forms a substantially denser ice–halite assemblage. If the exsolved salt remains mechanically coupled to the surrounding ice, this increase in density could reduce the buoyancy of the upwelling material and favour partial downward recycling. Alternatively, segregation of crystalline halite could promote the development of a \ce{NaCl}-rich layer near the transition. In either case, desalting would tend to inhibit, or at least reduce, the continued upward transport of salt towards the overlying ocean.

The desalting transition may therefore act as a pressure-controlled compositional filter, regulating not only how rapidly salt is transported but also how much of it can ultimately reach the surface ocean. Its efficiency will depend on the competition between convective entrainment, phase separation and gravitational recycling. Detailed numerical modelling will be required to determine what fraction of the exsolved \ce{NaCl} continues upwards and what fraction accumulates near the transition or is recycled into the deeper mantle.

Fig. \ref{int_struct_salt} illustrates how the potential importance of this filter varies with planetary mass and bulk ice fraction. As planetary mass increases, the $\sim5.1$\,GPa  level occurs at progressively shallower relative depth, and a larger fraction of the high-pressure ice mantle lies at pressures above this threshold. For massive water-rich planets ($> 2 M_E$), most of the high-pressure ice mantle may therefore lie within the pressure range in which NaCl-bearing ice VII is 
%\textcolor{ForestGreen}{
metastably retained at 300\,K. In this sense, a pressure-induced desalting transition could act as a compositional filter analogous to a differentiation boundary, controlling not simply the rate of salt transport but the fraction of salt that ultimately reaches the surface ocean. Quantifying this effect under realistic planetary geotherms will require constraining the temperature and concentration dependence of the desalting transition. 
%}

Several open questions remain. The concentrations studied here ($26.9$\,wt\%) far exceed realistic ocean salinities, but they could be representative of salt-rich brine zones developed at the ice/rock interface as a result of leaching processes \cite{Journaux2022}. Experiments at lower concentrations are needed to establish whether the substitutional regime and its volume expansion persist at various values of salt concentration. The high-temperature stability of the phase is unresolved as well. Frank \textit{et al.} observed halite exsolution above $500$\,K from low-concentration salty ice VII, but whether this applies to the structurally distinct substitutional phase is unknown. The behaviour of other geochemically relevant electrolytes, including \ce{KCl}, \ce{MgCl2} and \ce{CaCl2}, in ice VII also remains unexplored.

Despite these open questions, the present results demonstrate that high-pressure ice cannot always be treated as a passive, salt-excluding layer. The structural form in which salt is stored, and the pathway by which that structure forms, can substantially alter the density of the resulting assemblage and its response to decompression. Pathway-dependent incorporation and pressure-induced desalting may therefore provide a previously unrecognised control on chemical exchange between rocky interiors and subsurface oceans.

\section{Acknowledgements}
We acknowledge beamtimes HC-5468 at ID27 (ESRF), HC-5912 at ID15b (ESRF); BAG beamtime 20250410 at Psich\'e (Soleil); and beamtime 5-24-745 at D20 (ILL). 
We thank M. Mezouar for assistance at ID27; G. Garbarino and S. Gallego Parra for assistance at ID15b; N. Guignot for assistance at the Psich\'e beamline; and T. Hansen for assistance at the D20 beamline. 
L.E.B. acknowledges financial support from the European Union - NextGenerationEU (PRIN N. 2022NRBLPT). 
A.N., T.P., R.G. and L.E.B. acknowledge the financial support by the Swiss National Fund (SNF) under Grant No. 212889. 
L.E.B., G.T., R.C., and X.Z. acknowledge support from the ANR-23-CE30-0034 EXOTIC-ICE. 

\newpage

\bibliography{References}
 
\end{document}

\title{Supplementary Information: Pressure-induced desalting of NaCl-bearing ice VII: A compositional filter in water-rich planets}

\author{Alasdair Nicholls}
\email{alasdair.nicholls@epfl.ch}
\affiliation{Laboratory for Quantum Magnetism, Institute of Physics, \'Ecole Polytechnique F\'ed\'erale de Lausanne (EPFL), CH-1015 Lausanne, Switzerland}
\affiliation{Institut de Min\'eralogie, de Physique des Mat\'eriaux et de Cosmochimie (IMPMC), CNRS UMR7590, Sorbonne Universit\'e, Paris, 75252, France}
 
\author{Tomasz Por\k{e}ba}
\affiliation{Laboratory for Quantum Magnetism, Institute of Physics, \'Ecole Polytechnique F\'ed\'erale de Lausanne (EPFL), CH-1015 Lausanne, Switzerland}
\affiliation{Paul Scherrer Institute (PSI), CH-5232 Villigen PSI, Switzerland}
 
\author{Xi Zhu}
\affiliation{Universit\'e Paris Cit\'e, Institut de Physique du Globe de Paris (IPGP),  CNRS, Paris, 75005, France}
 
\author{Richard Gaal}
\affiliation{Laboratory for Quantum Magnetism, Institute of Physics, \'Ecole Polytechnique F\'ed\'erale de Lausanne (EPFL), CH-1015 Lausanne, Switzerland}

\author{Stefan Klotz}
\affiliation{Institut de Min\'eralogie, de Physique des Mat\'eriaux et de Cosmochimie (IMPMC), CNRS UMR7590, Sorbonne Universit\'e, Paris, 75252, France}
 
\author{Gabriel Tobie}
\affiliation{Laboratoire de Plan\'etologie et G\'eosciences, UMR 6112, Nantes Universit\'e, 2 Rue de la Houssini\`ere, Nantes 44322, France}

 \author{Henrik M. R{\o}nnow}
\affiliation{Laboratory for Quantum Magnetism, Institute of Physics, \'Ecole Polytechnique F\'ed\'erale de Lausanne (EPFL), CH-1015 Lausanne, Switzerland}

\author{Razvan Caracas}
\affiliation{Universit\'e Paris Cit\'e, Institut de Physique du Globe de Paris (IPGP),  CNRS, Paris, 75005, France}

\author{Livia Eleonora Bove}
\email{livia.bove@upmc.fr}
\affiliation{Laboratory for Quantum Magnetism, Institute of Physics, \'Ecole Polytechnique F\'ed\'erale de Lausanne (EPFL), CH-1015 Lausanne, Switzerland}
\affiliation{Institut de Min\'eralogie, de Physique des Mat\'eriaux et de Cosmochimie (IMPMC), CNRS UMR7590, Sorbonne Universit\'e, Paris, 75252, France}
\affiliation{Dipartimento di Fisica, Sapienza Universit\`a di Roma, Piazzale Aldo Moro 5, Roma, 00185, Italy}

\date{\today}

\maketitle

\section{Supplementary Results}

\subsection{Neutron Diffraction}

Additional experimental support for the concentration dependence of the expansion of the \ce{NaCl}-bearing ice VII was obtained from neutron diffraction measurements at the D20 beamline at the Institut Laue-Langevin (ILL), Grenoble, France. Neutron high-pressure measurements require the use of a Paris-Edinburgh (PE) Press which requires sample sizes in the order of $\sim$\,mm$^3$. Therefore, the sample was produced and loaded, as described in the main text. 
 
\begin{figure}
\centering
\includegraphics[width=0.95\columnwidth]{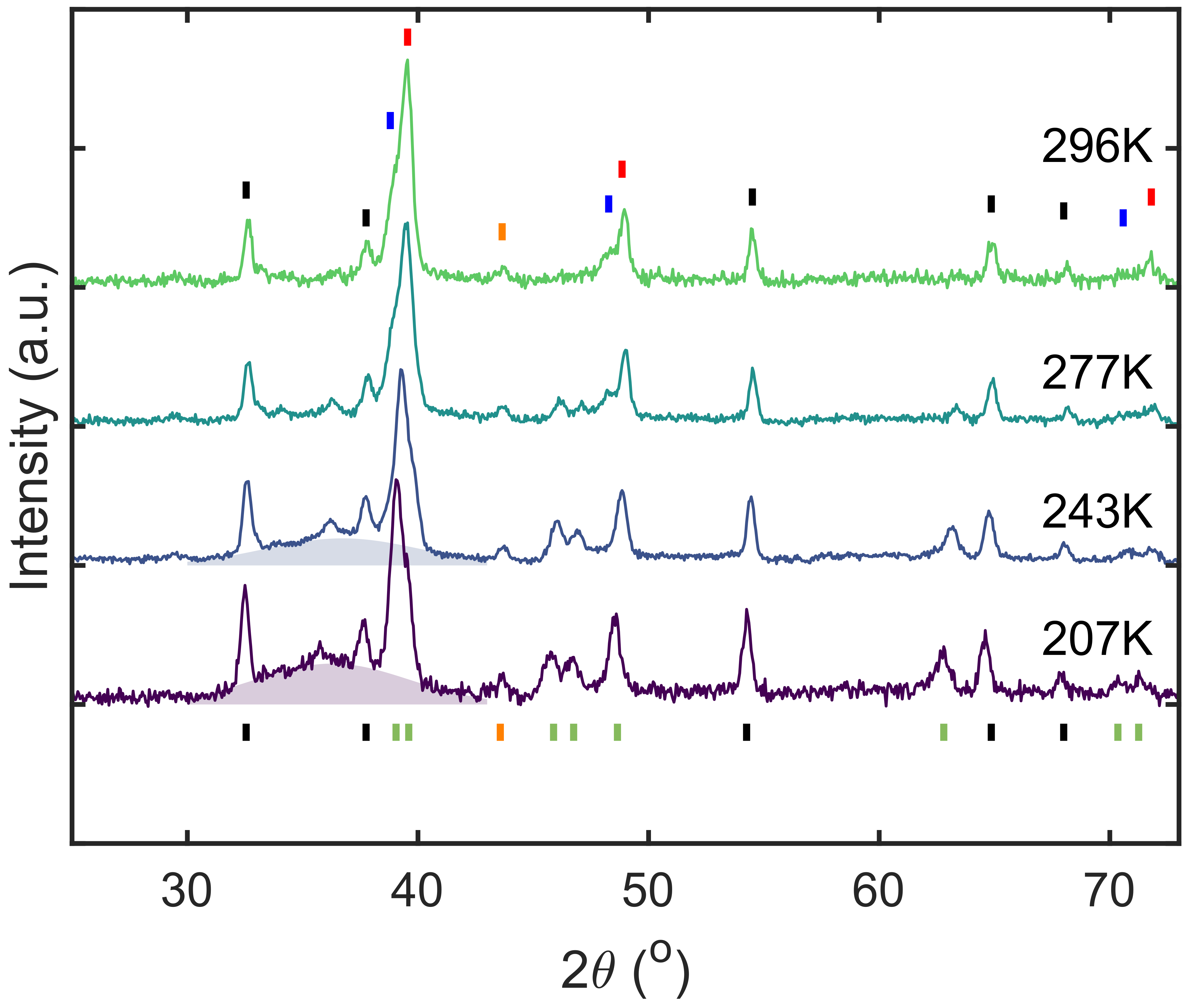}
\caption{Neutron Diffraction performed at D20, ILL as the \ce{NaCl}-bearing amorphous precursor was warmed up to room temperature at $P = 7.07$\,GPa. The wavelength was set at $\lambda = 1.54$\,\AA. Black tick marks indicate the Bragg peaks from Lead, the pressure calibrant; green tick marks are from pure ice VIII; red tick marks are from pure ice VII; blue tick marks are from the expanded \ce{NaCl}-bearing ice VII and finally, the orange tick mark is from the cBN anvils. The shaded region are indicating the amorphous bump which reduces as the temperature is warmed up until we see this expanded ice VII structure appear.}
\label{NeutronAmorph}
\end{figure}

The resulting amorphous material was cryoloaded into the PE press, compressed, and subsequently warmed to form crystalline ice VII (Supplementary Fig.~\ref{NeutronAmorph}). 

\begin{figure}
    \centering
    \includegraphics[width=0.95\linewidth]{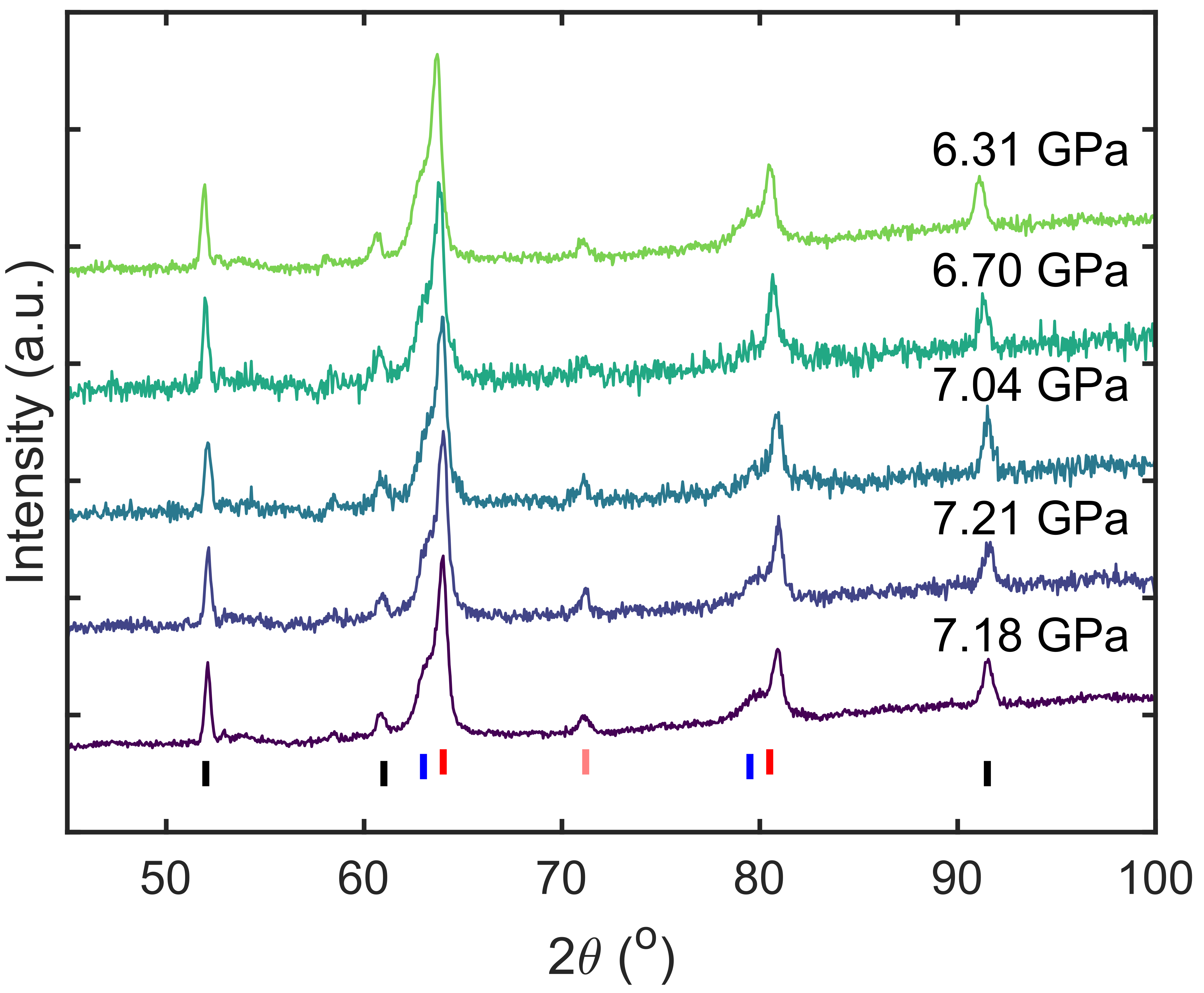}
    \caption{Neutron diffraction patterns of \ce{NaCl}-salty ice VII at the D20 beamline at ILL. The sample was formed following thermodynamic path No.1 in Fig.1 and temperature annealed at 7.18\,GPa. Red tick marks are from the pure ice VII phase, blue tick marks are from the expanded \ce{NaCl}-bearing ice VII, black tick marks are from the \ce{Pb} pressure calibrant and the orange tick mark is from the cBN anvils. The wavelength was  $\lambda = 2.41$\,\AA.}
    \label{Neutron}
\end{figure}

The resulting neutron diffraction patterns collected at room temperature and shown in Fig.~\ref{Neutron}, again reveal two cubic lattices. However, in this case, the expanded ice VII unit cell volume is greater than that of pure ice VII by only $\sim1$–$2$\,\%. This expansion is significantly smaller than that observed in the higher-concentration samples described in the main text.

During the spray-quenching process small, cubic crystalline inclusions were observed to have formed, but not collected, as their identity was uncertain at the time. Now that we see that there is a reduced lattice expansion measured with neutrons, it is plausible that these crystallites corresponded to crystalline \ce{NaCl} which had separated during the spray-quenching process. If partial desalting had occurred, the effective \ce{NaCl} concentration incorporated into the amorphous precursor, and thus into the ice VII lattice, would have been lower than the original solution concentration, resulting in a smaller expansion. Using Fig.\textcolor{blue}{7} in the main text, we can determine that the observed expanded ice VII with an expansion of $\sim1$–$2$\,\% corresponds to a salt concentration of $\sim5$\,wt\% of \ce{NaCl}. 

\subsubsection{Refinement Details}
Neutron diffraction data and X-diffraction data were treated with the \texttt{FullProf} Suite \cite{Rodrguez-Carvajal1993}. A Le Bail fitting scheme was utilised, to reliably obtain unit cell and peak-profile parameters for the Equation of State (Fig.\textcolor{blue}{5}) and desalting pressure (Fig.\textcolor{blue}{4}) analyses. Constrained Rietveld Refinements were used to determine the B$_{iso}$ values from the XRD data (Fig.\textcolor{blue}{2}a and b), and as reported in section III.A. Here, all the parameters of the refinement were refined one-by-one before B$_{iso}$ was refined last. Whilst refining B$_{iso}$, the peak-shape parameter (U) was held fixed to reduce correlations with B$_{iso}$. Although the value of B$_{iso}$ remained sensitive to parameter correlations, the substantially larger B$_{iso}$ of the \ce{NaCl}-bearing phase compared to pure ice VII was robust across the refinement strategies.

\newpage

\subsection{Room Temperature Compression Results}

\begin{figure}
    \centering
    \includegraphics[width=0.95\linewidth]{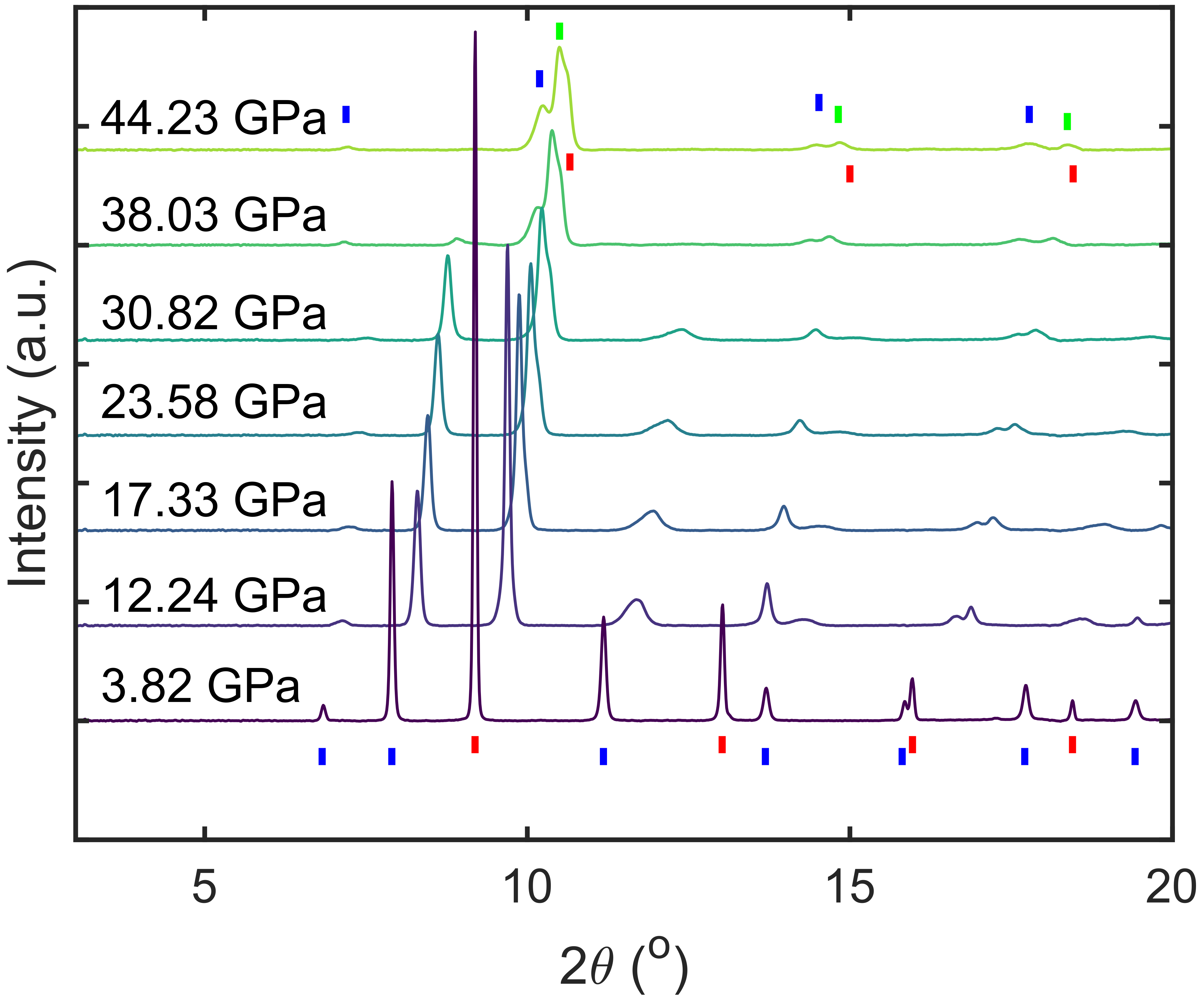}
    \caption{Evolution of the sample created from straight compression from the liquid to $3.82$\,GPa at $300$\,K. The sample was compressed up to $44.23$\,GPa. At the initial pressure, $3.82$\,GPa, pure \ce{NaCl}-B1 phase is indicated with the blue tick marks, whilst the pure ice VII peaks are indicated from the red tick marks. At the highest pressure, $44.23$\,GPa, the \ce{NaCl}-B2 phase is indicated with the blue tick marks, the pure ice VII by the red tick marks, and the slightly expanded ice VII by the green tick marks.}
    \label{Soleil_Evol}
\end{figure}

Final support for thermodynamic pathway dependence, and concentration dependent expansion of the \ce{NaCl}-bearing ice VII was obtained at the Psich\'e beamline at Soleil, Essonne, France. Here, a DAC was loaded with a salty aqueous solution with a \ce{NaCl} concentration of $26$\,wt\%. The cell was quickly compressed from the liquid phase and into the solid crystalline phase to $3.82$\,GPa. 
%\textcolor{ForestGreen}{
A pure ice VII phase, a slightly expanded ice VII and the pure crystalline \ce{NaCl} phase were observed via XRD in the centre of the cell at $44.23$\,GPa. The two peaks from the pure ice VII and slightly expanded ice VII were overlapping at low pressure, but due to the different bulk modulii of the phases, separated further at higher pressure. Fig.~\ref{Soleil_Evol} shows the evolution of the sample as it was compressed up to $44.23$\,GPa at $300$\,K.
%}

These results demonstrate that even upon rapid direct compression, without the `quench-and-compress' thermodynamic pathway, a small fraction of \ce{NaCl} can be incorporated into the ice VII lattice, producing a measurably expanded unit cell at higher pressures. However, the incorporation is incomplete. A large amount of crystalline \ce{NaCl} is clearly observed, and the expansion  of $\sim$3\,\%, corresponding to $\sim$8\,wt\% uptake (from Le Bail refinement) is substantially smaller than in samples prepared via the `quench-and-compress' pathway. This confirms that the thermodynamic pathway controls both the efficiency of incorporation and the resulting lattice expansion. We also note that the \ce{NaCl}-bearing ice VII peaks are significantly narrower than in the `quench-and-compress' prepared samples, consistent with reduced ionic disorder at lower effective salt concentrations — an observation that merits systematic investigation as a function of concentration. 

\newpage

\subsection{Thermal Correction of Calculations for \ce{NaCl}-bearing ice VII}

\begin{figure}
    \centering
    \includegraphics[width=0.95\linewidth]{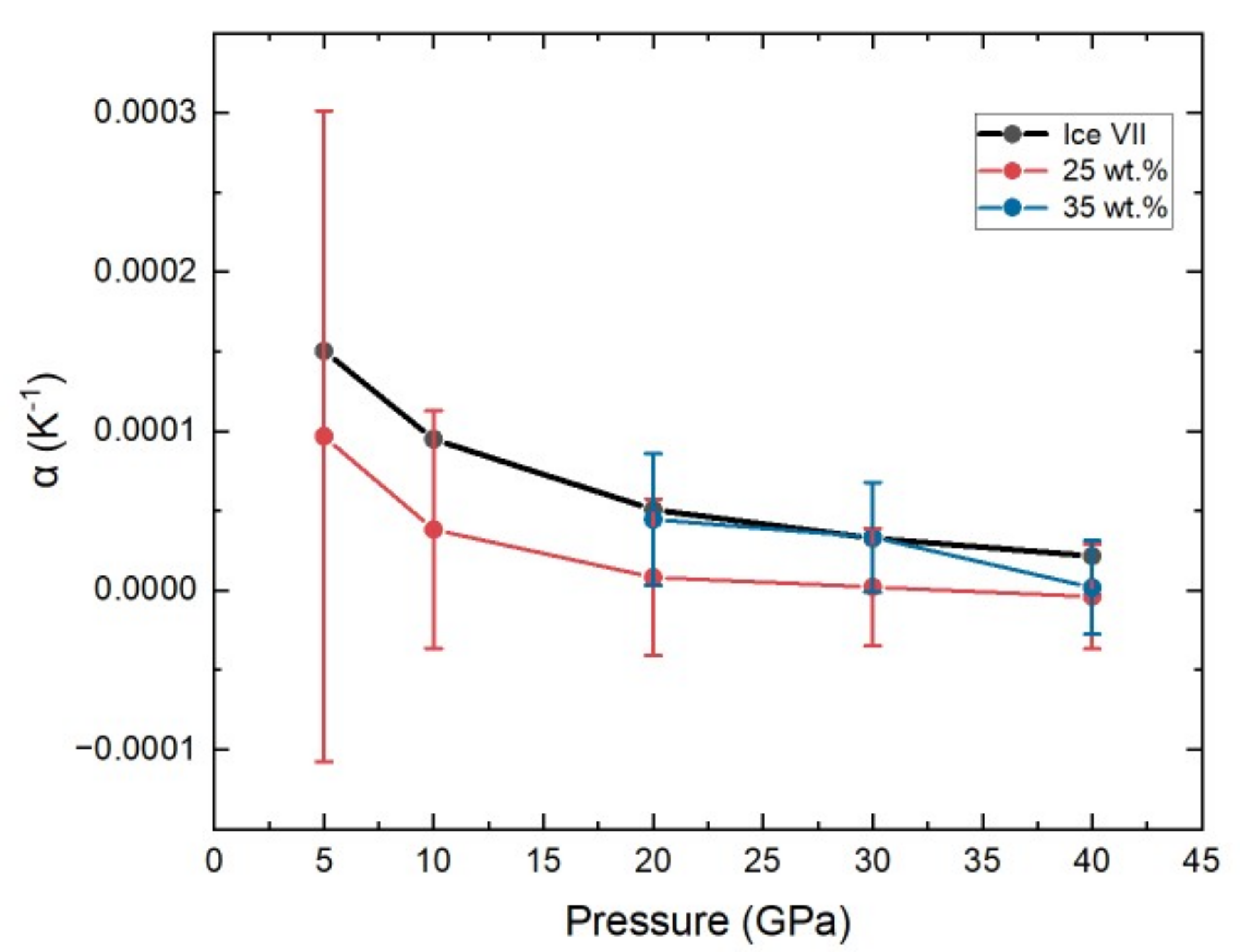}
    \caption{MD and static calculations of the impact of \ce{NaCl} salt on the thermal expansion $\alpha$ of ice VII.  Static calculations were performed for \ce{NaCl} contents of 0, 6, 19, and 35\,wt\%, and the resulting volumes were fitted as a function of salt concentration. The fitted static volume--composition relation was then compared with the MD results at 0, 19, and 35\,wt\% \ce{NaCl} to calculate the thermal expansion.
}
    \label{Alpha_Correction}
\end{figure}

The \textit{ab initio} calculations were performed for \ce{NaCl}-bearing ice VII at 0\,K. The volumes were subsequently scaled up to $300$\,K using the thermal expansion of pure ice VII. This can be justified since MD and static calculations indicated that the impact of \ce{NaCl} salt did not have any incorporation on the thermal expansion of ice VII, as shown in Fig.~\ref{Alpha_Correction}.

\newpage

\bibliography{References}